# The Microsoft TerraServer™

Tom Barclay, Robert Eberl,
Jim Gray, John Nordlinger, Guru Raghavendran,
Don Slutz, Greg Smith, Phil Smoot
Microsoft Research and Development

John Hoffman, Natt Robb III,
Aerial Images

Hedy Rossmeissl, Beth Duff, George Lee, Theresa Mathesmier, Randall Sunne
United States Geological Survey

Lee Ann Stivers, Ken Goodman
Digital Equipment Corporation

# Microsoft TerraServer™

Tom Barclay, Robert Eberl,
Jim Gray, John Nordlinger, Guru Raghavendran,
Don Slutz, Greg Smith, Phil Smoot
Microsoft Research and Development

John Hoffman, Natt Robb III,
Aerial Images

Hedy Rossmeissl, Beth Duff, George Lee,
Theresa Mathesmier, Randall Sunne
United States Geological Survey

Lee Ann Stivers, Ken Goodman
Compaq Corporation

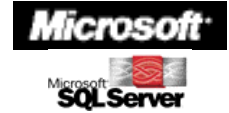

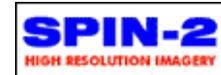

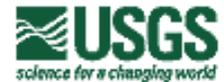

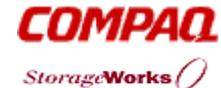

June 1998

**Abstract**
The Microsoft TerraServer stores aerial and satellite images of the earth in a SQL Server Database served to the public via the Internet.  It is the world's largest atlas, combining five terabytes of image data from the United States Geodetic Survey, Sovinformsputnik, and Encarta Virtual Globe™. Internet browsers provide intuitive spatial and gazetteer interfaces to the data.  The TerraServer demonstrates the scalability of Microsoft's Windows NT Server and SQL Server running on Compaq AlphaServer 8400 and StorageWorks™  hardware. The TerraServer is also an E-Commerce application.  Users can buy the right to use the imagery using Microsoft Site Servers managed by the USGS and Aerial Images.  This paper describes the TerraServer's design and implementation.

# Table of Contents

# The Microsoft TerraServer™

The TerraServer has five terabytes of satellite and aerial images of urban areas compressed to one terabyte of database data. It serves these images onto the Internet with a graphical and intuitive user interface. The application demonstrates several things:
- **Information at your fingertips:** This is the most comprehensive world atlas anywhere — and it is available to anyone with access to the Internet.
- **Windows NT Server and SQL Server can scale up to huge nodes**: The TerraServer fills eight large cabinets: one for the Compaq Alpha 8400 processors, and seven cabinets for the 324 disks -- almost three terabytes of raw disk storage and 2.3 TB of RAID5 storage.
- **Windows NT and SQL Server are excellent for serving multi-media and spatial data onto the Internet**.
- **Microsoft Site Server can help sell images over the Internet.**

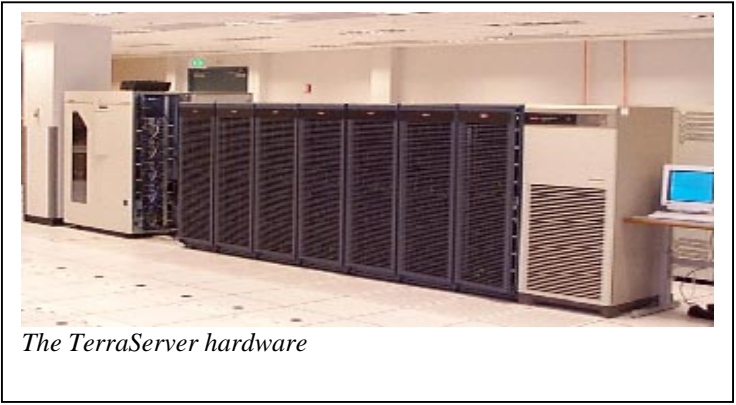
*The TerraServer hardware*

TerraServer is a multi-media database that stores both classical text and numeric data, as well as multi-media image data. In the future, most huge databases will be comprised primarily of document and image data. The relational meta-data is a relatively small part of the total database size. TerraServer is a good example of this new breed of multi-media databases.

## The Application

**An Interesting Internet Server**: TerraServer is designed to be compelling Internet application. It tries to be interesting to almost everyone everywhere, be offensive to no one, and be relatively inexpensive to build and operate. It is hard to find data like that — especially a terabyte of such data. A terabyte is nearly a billion pages of text — 4 million books. A terabyte holds 250 full-length movies. It is a lot of data.

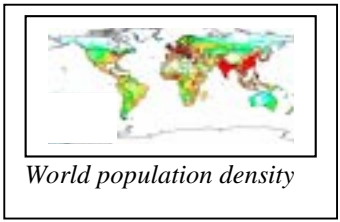
*World population density*

**Satellite Images of the Urban World:** Pictures are big and have a universal appeal, so it was natural to pick a graphical application. Aerial images of the urban world seemed to be a good application. The earth's surface is about 500 square tera-meters. 75% is water, 20% of the rest is above 70° latitude. This leaves about 100 square tera-meters. Most of that is desert, mountains, or farmland. Less than 4% of the land is urban. The TerraServer primarily stores images of urban areas. Right now, it has nearly five square tera-meters -- and it grows as more data becomes available.

**Cooperating with United States Geological Survey**: The USGS has published aerial imagery of many parts of the United States. These images are approximately one-meter resolution (each pixel covers one square meter.) We have a Cooperative Research Agreement (CRADA) with the USGS to make this data available to the public. We have loaded all the published USGS data (3 TB raw, 0.6 TB compressed). This is 30% of the United States. As additional data becomes available, it will be loaded into the TerraServer. This data is unencumbered and can be freely distributed to anyone. It is a wonderful resource for researchers, urban planners, and students. The picture at left shows a baseball game in progress near San Francisco. You can see the cars, but one-meter resolution is too coarse to show people.

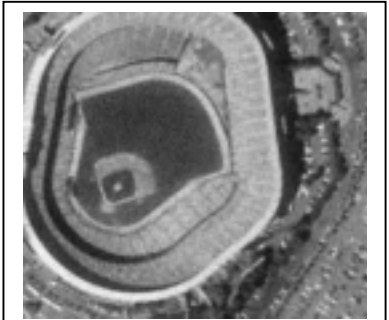
*A USGS 1-meter resolution image of Candlestick Park near San Francisco*

**Working with Sovinformsputnik (the Russian Space Agency) and Aerial Images.** To be interesting to everyone everywhere, TerraServer must have worldwide coverage. The USGS data covers much of the continental United States. There is considerable imagery of the planet, but much of it either has poor quality (10 meter to 1-km resolution), or has not been digitized, or is encumbered. Sovinformsputnik and their representative Aerial Images have some of the best data and were eager to cooperate. The Russians and Aerial Images contributed two square tera-meters of imagery (1.56-meter resolution). This data is trademarked SPIN-2, meaning satellite-2-meter imagery. They intend to deliver an additional 2.4 square tera-meters over the next year.

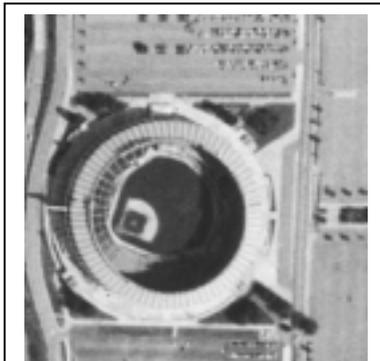
*A SPIN-2 1.6-meter image of Atlanta's Olympic stadium.*

**TerraServer is the largest world atlas**: The Sovinformsputnik SPIN-2 imagery covers Rome, Athens, Hong Kong, New York, Chicago, Seattle, and many other cities. TerraServer has more data in it than ALL the HTML pages on the Internet. If printed in a paper atlas, with 500 pages per volume, it would be a collection of 2,000 volumes. It grows by 10,000 pages per month. Clearly, this atlas must be stored online. The USGS data (the three square tera-meters) is seven times larger. This data is a world-asset that will likely change the way geography is taught in schools, the way maps are published, and the way we think about our planet.

**TerraServer as a business.** Slicing, dicing, and loading the SPIN-2 and USGS data is a continuing process. Today, the TerraServer stores a terabyte. Aerial Images, Compaq, and Microsoft are operating the TerraServer on the Internet (www.microsoft.com/ TerraServer). Microsoft views TerraServer as a demonstration for the scalability of Windows NT Server and Microsoft SQL Server. Compaq views it as a demonstration for their Alpha and StorageWorks servers. The USGS is participating as an experiment on presenting USGS data to a wider audience through the Internet. They operate an online store that allows anyone to download copies of the USGS images. Sovinformsputnik and Aerial Images view TerraServer as a try-and-buy distributor for their intellectual property. They make coarse-resolution (8-meter, 16-meter, and 32-meter) imagery freely available. The fine-resolution data is viewable in small quantities, but customers must buy the right to use the "good" imagery. All the SPIN-2 images are watermarked, and the high-resolution images are lightly encrypted.

**Site Server, a new business model for the Internet.** Aerial Images' business model is likely to become a textbook case of Internet commerce. By using the Internet to sample and distribute their images, Aerial Images has very low distribution costs. This allows them to sell imagery in small quantities and large volumes. Microsoft helped USGS and Aerial Images set up Microsoft Site Servers that accept credit-card payments for the imagery. A "download" button on the image page takes the user to these Site Servers (Microsoft has no financial interest in these transactions). You can buy a detailed image of your neighborhood for a few dollars.

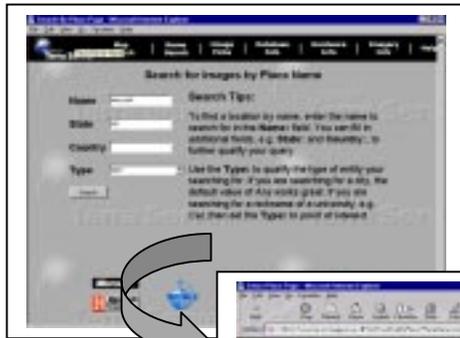
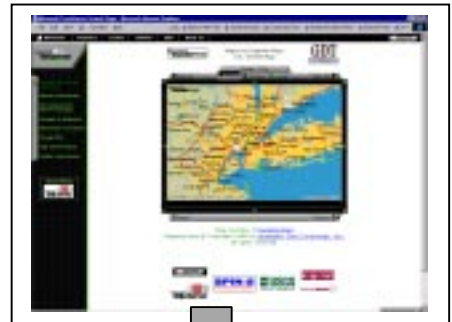

*Navigation can be via name (left) or via a map (right). In either case, the user can select the USGS or SPIN-2 images for the place*

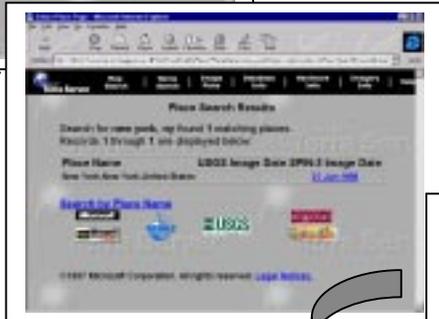

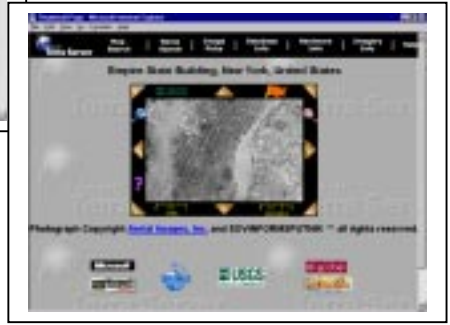

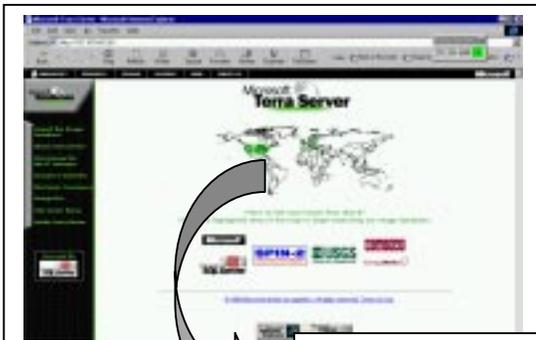

*The coverage map at left shows green where the TerraServer has imagery. By clicking on the coverage map, users can quickly zoom on a particular spot.*

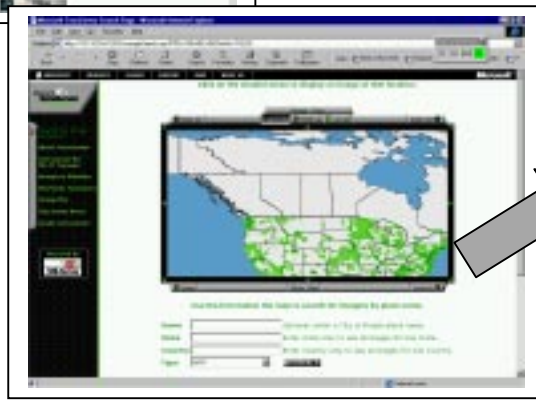

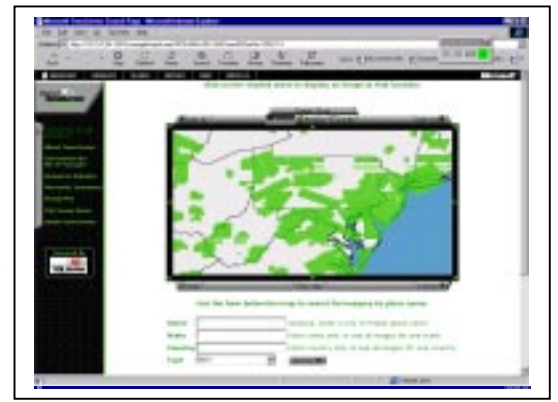

## User Interface to the Microsoft TerraServer

**Navigation via database searches:** The TerraServer can be accessed from any web browser (e.g. Internet Explorer, Netscape Navigator). Full resolution SPIN-2 imagery requires the web browser to support Java™ applets. Any web browser that supports HTML tables and display of Jpeg data can host the TerraServer user interface. Navigation can be a spatial via a point-and-click map control based on Microsoft's *Encarta® World Atlas.* Clients only knowing the place name can navigate textually by presenting a name to the *Encarta Virtual Globe Gazetteer*. The gazetteer knows the names and locations of 1.1 million places in the world. For example, "Moscow" finds 28 cities, while "North Pole" finds 5 cities, a mining district, a lake, and a point-of-interest. There are 378 San Francisco's in the Gazetteer. The user can select the appropriate member from the list. The map control displays the 40-km map of that area. The user can then pan and zoom with this map application, and can select the USGS and SPIN-2 images for the displayed area.

**Navigation via *coverage map*.** The USGS gave us a shaded relief map (Mercator projection) which includes political (state and province) boundaries. We shaded this map green where we have some imagery. Then we built an image pyramid: one image for the whole planet, and two-levels of zoom in that covers a continent and then a region. The bottom panels of the previous page show an example of zooming in on New York City. We added this interface last, but it is the most popular way to navigate the database.

**Spatial navigation via the map control**. A dynamic HTML page allows the client browser and to talk with the Microsoft Expedia™ map server (http://www.expediamaps.com/) that provides the basic features of Microsoft *Encarta World Atlas*™ and Microsoft *Automap Streets*™ as GIF images. The applet lets users pan and zoom over graphic images of the earth and of US street maps. The applet decides what the client wants to see and sends a request for that map to the Expedia map server at http://maps.expedia.msn.com/. That server, given the corners of a rectangle and an altitude, generates the view of the earth inside that rectangle. It generates a GIF image that is downloaded to the applet in the client browser. The map server is provided by MSN to any Internet customer. We have just wrapped it in our Java applet. This app works on Windows, Macintosh, and UNIX clients. Coverage map and spatial access is especially convenient for those who do not understand English.

**Zooming in and out:** The map controls allows the browser to zoom out and see a larger area, or zoom in and see finer detail. The coarsest view shows the whole planet. The user can "spin" the globe to see the "other side" and place the point of interest in the center of the screen. Then the user can zoom in to see fine detail. Where we have street maps (*Microsoft Automap® Streets*), the zoom can go all the way down to a neighborhood.

*Moving Around: Once you find the spot you are looking for, you can see nearby places by pushing navigation buttons to pan and zoom. Doing this, you can "drive" cross-country.*

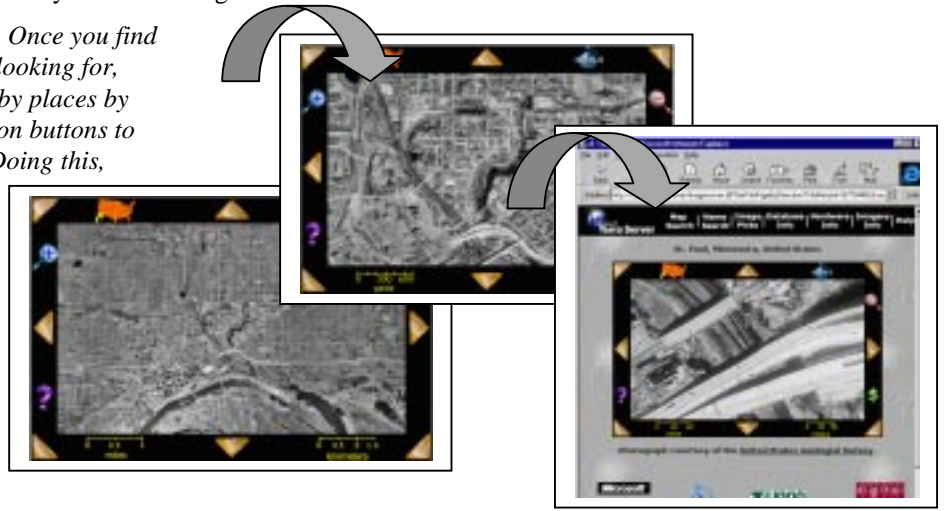

**Encarta, USGS, and SPIN-2 Themes:** TerraServer has several different views of the earth: the coverage map view, the *Encarta World Atla*s view, the USGS image view, and the Sovinformsputnik-SPIN-2 view. We call each of these a *theme*. The user may switch from one theme to another: perhaps starting with the Encarta theme, then the SPIN-2 theme, and then the USGS theme of the same spot. With time, we expect to have multiple images of the same spot. Then the user will be able to see each image in turn. Your grandchildren will be able to see how your neighborhood evolved since 1990.

**Download**: If you like the SPIN-2 imagery or USGS image, you can push the "DownLoad Image" button. That takes you to the Aerial Images or USGS Site Server. Both Aerial Images and USGS E-commerce sites run Microsoft's Commerce Server™, which is a component of Microsoft Site Server. That is where the similarity between Aerial Images and the USGS e-commerce site ends. The Aerial Images web site allows Internet users to select one of three sizes for a digital image. The user can also select a choice of format (TIF or JPEG) and can have also have a photograph printed and delivered over-night by Kodak.

The USGS site allows users to download the image viewed on Microsoft TerraServer as a single digital image in JPEG format at no charge. The USGS site offers an easy method to purchase one or more of the original data sets used to form these images. The original USGS imagery is intended for use by professional GIS users. The Site Server allows you to shop for imagery and quotes you a price. If you want to purchase a high-quality digital copy of these images, Site Server asks for your credit card, debits it, and downloads the images you purchased or ships them to you on the media of your choice.

The USGS and SPIN2 sites are a good example of selling soft goods over the Internet.

*The Microsoft TerraServer* 5

## Server Design

The TerraServer has several components that combine to make a seamless Internet application.

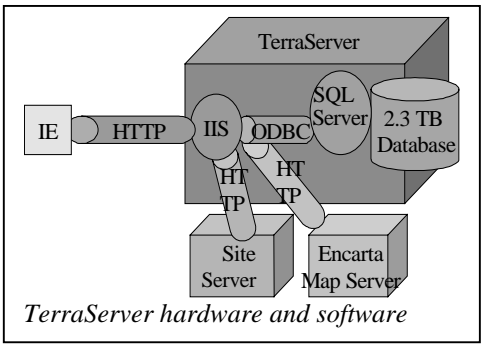

*TerraServer hardware and software*

**Internet Information Server and Active Server Pages:** Clients send requests to the TerraServer's Internet Information Server (IIS) built into Windows NT. These requests are passed to Active Server Pages (ASPs) programmed in VBscript. These ASPs, programmed in VBscript send queries to stored procedures in the SQL Server database to query the Gazetteer and to fetch image tiles. The ASPs dynamically construct the HTML web pages needed to mosaic the tiles together to make a complete image. The server first returns the HTML for the outer frame, and the HTML table referencing the two dimensional array of tiles. The number of tiles display in the HTML table is controlled by the user and the current page's image resolution. TerraServer stores imagery in 32m/p (square meters/pixel), 16 m/p, 8m/p, and full resolution (1m/p or 1.56m/p). The user decides if they want to see a small, medium, or large view. A full resolution "small" view page display 4 image tiles whereas a "large" view page displays 16 image tiles. A 32m/p web pages range from 64 tiles in a "small" view to 256 image tiles in a "large" view. When the client pans an image, 50% of the images on the current frame are moved to another location within the frame and the other 50% are downloaded from the database.

The VBscript program dynamically creates the necessary HTML to render that image. It sends this HTML back to the client's browser. The client browser then requests the images needed to fill in the picture. Depending on the image size the user selects, this can be between 4 and 256 tiles. These URL requests generate between 30 and 500 database accesses.

**Tiled Image Database:** The database stores both the USGS data and the SPIN2 data as small (10 kilobyte or less) tiles compressed with JPEG. Larger images are created as a mosaic of these tiles. This allows quick response to users over slow voice-grade phone lines. It also allows the application to pan and zoom across the images.

**Gazetteer**: The *Encarta World Atlas Gazetteer* has over a million entries describing most places on earth. All these records are stored and indexed by Microsoft SQL Server. Stored procedures to look up these names and produce an HTML page describing the top 10 "hits", with hot-links to the images if they are in the TerraServer.

**TerraServer uses SQL Server 7.0:** TerraServer uses the 1998 version of Microsoft SQL Server. This version supports larger page sizes, has better support for multi-media, supports parallelism within queries, parallel load, backup and restore utilities, and supports much larger databases. TerraServer has been a good stress test for SQL Server 7.0.

**Loading the Database:** SOVINFORMSPUTNIK and the USGS delivered the data to us on several hundred tapes. We had to sort, reformat, slice, and dice this data before it could be inserted in to the database. We wrote several programs to do this image processing. We also wrote a load manager that consumes these files and feeds the data into the TerraServer using the SQL Server loader APIs (ODBC BCP). Using several parallel streams, it loads at approximately 2 MBps. At this rate, the load takes 6 days. The load is more constrained by the scan, slice, and dice process than by the SQL load rate. Indeed, the database load rate is 15 MBps, eight times faster than the load program can produce the data.

**Site Server**: If a client wants to buy some Imagery from Aerial Images, the client pushes the Down Load Image button. Site server uses secure HTTP to authenticate the user, quote the user a price for the requested data. The data providers, Aerial Images and the USGS, have built electronic "stores" tailored to their existing products and their markets. The sites differ in pricing philosophies and market focus. The Aerial Images site is designed for unsophisticated users. Imagery size and formats are designed to be attractive to non-professional users interested in a digital photo of a small area. The USGS site on the other hand, is targeted towards the USGS' traditional market of GIS professionals and data resellers. Both systems use credit cards as authorization for payment. Aerial Images' site downloads digital images immediately over the user's internet connection. Paper photographs are delivered via ground mail or overnight express services. USGS images are distributed via ground mail carriers only.

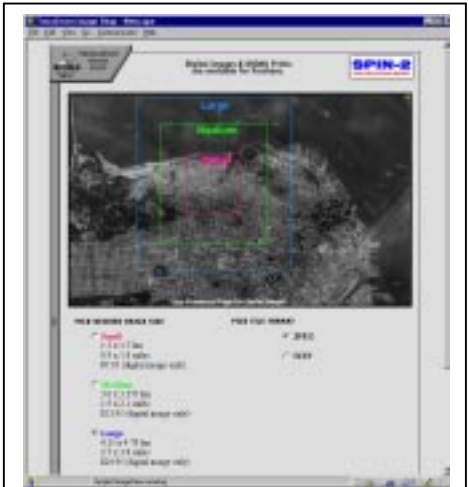

*Customers can buy the right to use SPIN2 images. They may also buy a Kodak print of the image. Larger images are more expensive. USGS offers similar services for its data. All this commerce is performed by Microsoft Site Servers operated by SPIN2 and USGS.*

The diagram below describes how Microsoft TerraServer integrates with the E-Commerce web sites operated by the USGS and SPIN-2 organizations. Web users access Microsoft TerraServer through their PC connected to the Internet. Microsoft TerraServer web application is built with IIS 4.0 Active Server Pages. The Active Server Pages access the TerraServer SQL database to built HTML pages and references to image objects contained in the TerraServer Sphinx database. The HTML pages are sent over the internet to the web user's PC. The "Download" button on each Microsoft TerraServer image page opens a URL document at the data provider commerce site.

The USGS electronic commerce site runs on Microsoft Commerce Server on a computer system located at the USGS EROS Data Center in Sioux Falls, South Dakota. The Microsoft Commerce Server application was written by the USGS staff members and integrates into the USGS' automated order entry, inventory, and shipping systems. The Microsoft Commerce Server handles the creation of the order and all payment reconciliation between the web user and the USGS' bank institution. The USGS site is secured using SSL technology and certificates obtained through VeriSign.

**The USGS electronic Store:** The USGS electronic store is designed for professional image users. The USGS distributes Microsoft TerraServer images to web users at no charge. These images are "Jpeg compressed" images that do not have the image quality required for certain professional applications. The USGS electronic store enables professional users to purchase the same uncompressed imagery used to create the Jpeg images stored in the Microsoft TerraServer database. Because of the size of the uncompressed imagery, which is typically 46MB per file, the USGS distributes uncompressed imagery on CDROM media.

**The SPIN-2 Commerce Server:** The SPIN-2 commerce site runs on Microsoft Commerce Server on a DEC Alpha 800 server located at Aerial Images, Inc. in Raleigh, North Carolina. The SPIN-2 electronic commerce site is intended to be used by either professional or non-professional users requiring images that are physically larger than images they can view on Microsoft TerraServer. At the SPIN-2 store, users can select the physical size of the image, the type of image format, and output media. In addition to a digital image, users can opt for a photographic print of their selected image in one of three sizes ranging from 8 ½ x 11" to 20" x 26" prints. SPIN-2 will forward your digital image to Kodak for processing.

**Commerce Server talks to the banks:** The Microsoft Commerce Server application steps the user through the selection process, obtains their payment information, and connects to SPIN-2's bank through PayLinks, a third-party payment component that integrates with Microsoft Commerce

Server. Once the credit card information is validated, the Commerce application inserts an order into the SPIN-2 "TerraShop Database".

**How TerraShop delivers data to the customer.** A custom program picks up the order from the TerraShop database, constructs the image requested by the user and routes the image to the appropriate locations. The digital image is output to a "customer pickup area" on the SPIN-2 web site. The purchaser is sent an e-mail message notifying them that their order is ready to pickup. If a paper print was ordered, the digital image is electronically placed in the "Kodak pick up area". At regular intervals, Kodak's automated operation picks orders up for printing and delivery the same day.

**Four web sites working together:** The Microsoft TerraServer is not only a huge database but it is also an example of how to build a distributed electronic store. Four separate web sites located at opposite ends of the United States seamlessly interoperate to provide interactive shopping, payment processing, and product delivery

**Summary**: TerraServer is a new world atlas — far larger than any seen before. It is a relatively simple database application, but it demonstrates how to build a real Internet application using WindowsNT and SQL Server running on Compaq AlphaServer 8400 and StorageWorks servers.

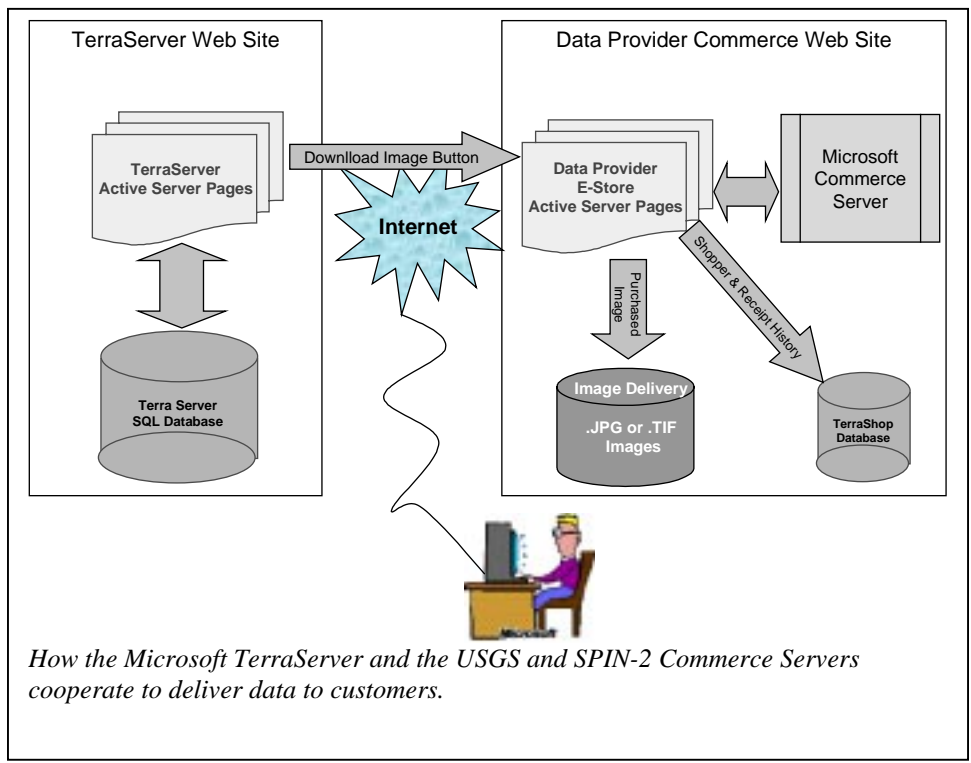

*How the Microsoft TerraServer and the USGS and SPIN-2 Commerce Servers cooperate to deliver data to customers.*

## Database Design

The TerraServer presents an interesting geo-spatial database design problem. It contains data from three different sources represented in different coordinate systems. It has to integrate all this data into a single intuitive user interface. This section describes how the data is represented in the database and how it is indexed.

### Coordinate Systems

The earth is not flat. It is not round either -- it is a bumpy oblate spheroid. When measuring the earth at one-meter resolution, this becomes a very important issue.

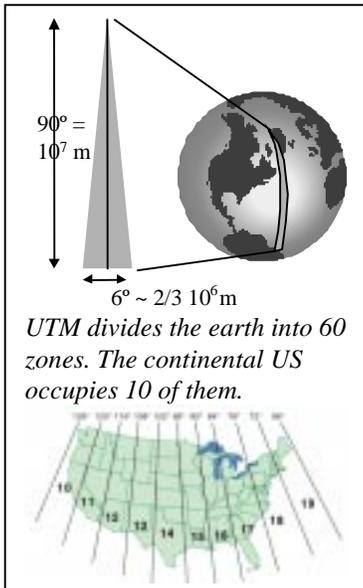

*UTM divides the earth into 60 zones. The continental US occupies 10 of them.*

**USGS DOQs:** The USGS has aerial photographs of most of the United States. It has corrected these aerial photographs for elevation and camera optics. The USGS then maps the true image into the Universal Transverse Mercator (UTM) coordinate system. The resulting digital images are *mosaiced* into Digital Orthorectified Quarter Quadrangles (DOQQs for short). A quadrangle is a one-eighth degree square (about 3.5 kilometers on a side). The USGS has published paper quadrangle maps for many decades. A DOQQ is one-quarter of a USGS quadrangle.

**The UTM system:** UTM divides the earth into 60 zones. Each zone is two 6º spherical triangles going from the equator to the poles. The continental United States occupies nine UTM zones (Alaska and Hawaii add 7 more zones). A UTM projection flattens each of these spherical triangles (projects them onto a plane). The meridian of the triangle is represented perfectly, but all the other pixel-points are slightly distorted to be trapezoids rather than squares. In particular, the pixels at the edges of a zone have north at 3º from the vertical. The UTM system maps latitude lines into curves. This is barely noticeable to the eye, but is very noticeable when images that lie on zone boundaries are concatenated.

**USGS data uses UTM**: We decided to use the USGS coordinate system for the USGS data. To be exact, the USGS uses UTM with the NAD83 datum. It would be too much work for us to remap the USGS data into a coordinate system that gives a seamless mosaic of the earth. In UTM, each point has a zone number, then a Northing (meters from the equator), and an Easting (meters from the west meridian of the triangle). TerraServer USGS images are a fixed size – 1800 meters by 1200 meters. The TerraServer assigns a unique UGridID to each TerraServer image by concatenating the UTM zone with the image's Easting ID (Easting + 400 / 1800) number followed by a bit interleave of the Northing ID (Northing / 1200). The bit interleaving causes nearby images to have a common UGridID prefix.

**SPIN-2 uses latitude-longitude:** The SPIN2 data is taken from 200 kilometers up. An original SPIN-2 image is a 40-km by 160-km photographic swath taken by a former Russian military satellite. These are declassified photos. A recent US-Russian treaty allows Russia to export to the United States. Each photo has a resolution of 1.56 square meters per pixel. We have 2 trillion square meters of these images (0.7 TB). Each 40km by 160km photograph is scanned into four separate 40km X 40km image because it is too large to scan all at one. The digital scan is separated into four separate 20km X 20km files because Adobe Photoshop cannot rotate an image larger than 30,000 pixels. Aerial Images personnel geo-locate five points on one 20km by 20km image. Photoshop is used to rotate each 20km by 20km quadrant. This creates the appearance of a diamond shaped photo within a square white canvas. The upper left corner point and lower right corner point of the square image is computed. The upper left and lower right corner points of the other three images are computed relative the first image. Our image editing program reads each "Spin-2" 20km X 20km image and creates "TerraServer SPIN-2" images that are $1/48^{th}$ of degree wide by $1/96^{th}$ of a degree high. Pixels from adjacent 20km X 20km image are merged creating a single "TerraServer" Spin-2 image. We mapped the SPIN2 data into a latitude-longitude reference system -- each image is given a unique Z-Grid ID, which is the interleaving of the latitude and longitude of the center point of the TerraServer SPIN-2 image. On earth, there are a total of 298598400 unique ZgridID values. ((360 longitude degrees * 48 'cuts per degree') x (180 latitude degrees x 96 'cuts per degree'))

### Image Format

**Images are 256-level gray scale JPEG:** Browsers generally reduce the number of levels displayed. The images are all stored JPEG compressed to 80% be faithful to the original image. This typically gives a 5:1 compression.

**The image tile pyramid:** The images are stored in the database as an image pyramid so that users can zoom in and out (see Figure 1). An additional constraint is that no image should be much larger than 10KB. This constraint comes from the need to support clients accessing the database via 28.8 kbps modems. It takes about 3 seconds to download 10KB image. A complete web page is made as mosaics of these small images.

**Tile, thumb, browse, and jump images:** Large images are first sliced into tiles that are about 10 kilobytes each. Each of these tiles covers a tiny area (less than a tenth of a square kilometer). The TerraServer returns a mosaic of these tiles to the user on each query. Coarser tile resolutions are stored to support zooming. The data load process mosaics tiles and then dithers them down to wider-panorama images. For USGS data, an 8x8 mosaic is dithered down to 8-meter resolution produces a *browse image* from the original images. For the SPIN2 1.56-meter data, 5x5 array of tiles is dithered to form the *browse image*. These browse images are further dithered down to 16-meter (thumbnail image) and 32-meter (jump image) resolution images. Because these images have lower resolution, they occupy $1/64^{th}$, $1/256^{th}$, and $1/1024^{th}$ of the space of the tile images. That is, they occupy almost no space at all.

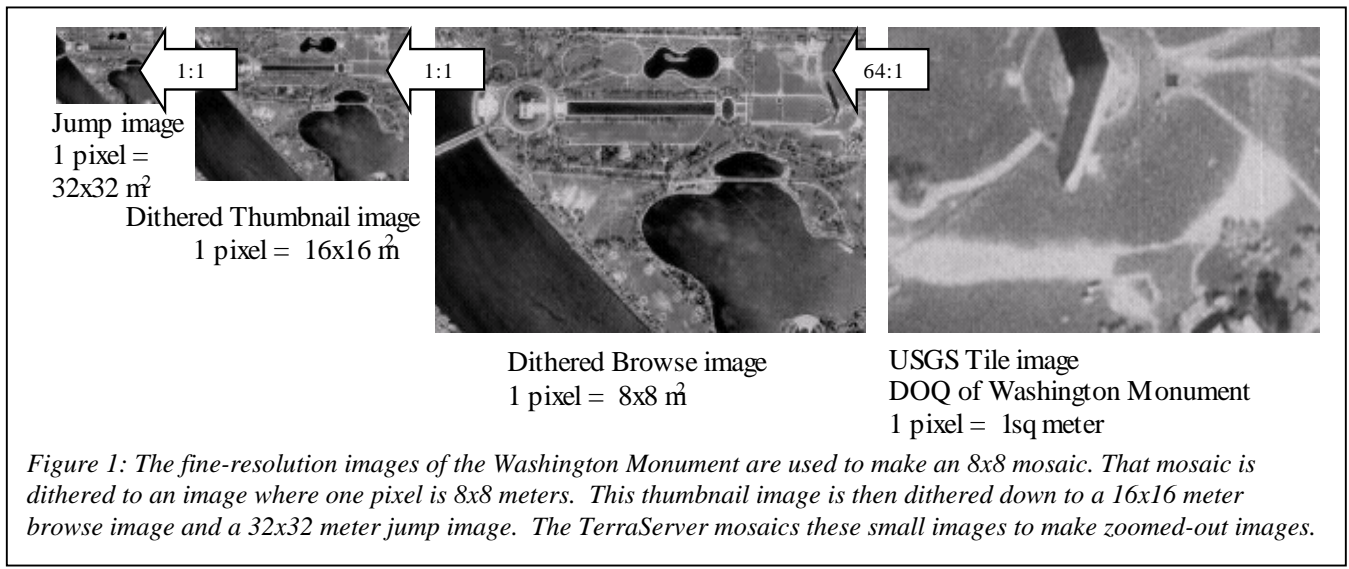

*Figure 1: The fine-resolution images of the Washington Monument are used to make an 8x8 mosaic. That mosaic is dithered to an image where one pixel is 8x8 meters. This thumbnail image is then dithered down to a 16x16 meter browse image and a 32x32 meter jump image. The TerraServer mosaics these small images to make zoomed-out images.*

**Tile 4x3 aspect ratio:** Tile sizes were chosen to approximately match the 4x3 computer display aspect ratio. The USGS data follows the USGS coordinate system. In the USGS data set, each tile is exactly 225x150 meters. Based on visual experiments, we chose a browse image size of $1/48^{th}$ of a degree wide and $1/96^{th}$ of a degree high. This translates to an area about 1 km x 2km. Further north, the $1/48^{th}$ of a degree wide shrinks to almost nothing. In the middle latitudes, it has the desired 3x4 ratio. Each thumbnail is sliced 5x5 to make the tile images.

Once a tile size is chosen, all the other sizes are a pyramid derived from that basic unit. Table 1 gives the approximate sizes and cardinalities of the data sets.

| **Table 1:** Cardinalities and sizes of the USGS and SPIN2 data sets as stored in the TerraServer. The SPIN2 data tiles vary in size because the thumb images are $1/48^{th}$ x $1/96^{th}$ of a degree. | | | | |
|---|---|---|---|---|
| | Jump | Browse | Thumb | Tile |
| SPIN2 | | | | |
| Resolution per Pixel | 32 meter | 16 meter | 8 meter | 1.6 meter |
| Pixels | 44x34 | 88x68 | 176x134 | 167x131 to 239x152 |
| Area | ~ 5 km$^2$ | ~ 5 km$^2$ | ~ 5 km$^2$ | ~ .1 km$^2$ |
| Image Size (bytes) | ~ .4 KB | ~ 1.5 KB | ~ 6 KB | ~ 6 KB |
| Cardinality | 650 k | 650 k | 650 k | 16 m |
| GigaBytes | .24 GB | 1 GB | 4 GB | 96 GB encrypted |
| USGS DOQs | | | | |
| Resolution | 32 meter | 16 meter | 8 meter | 1 meter |
| Pixels | 56x37 | 112x75 | 225x150 | 225x150 |
| Area | 2.1 km$^2$ | 2.1 km$^2$ | 2.1 km$^2$ | .03 km$^2$ |
| Image Size (bytes) | ~ .4 KB | ~ 1.7 KB | ~ 7 KB | ~ 7 KB |
| Cardinality | 1.5 m | 1.5 m | 1.5 m | 96 m |
| GigaBytes | .6 GB | 2.4 GB | 10 GB | 700 GB |

### Database Themes

As explained so far, there are two separate data themes: USGS DOQs and SPIN-2 satellite images. Each theme has it's own set of SQL tables. Each image along with its meta-data is a record in the database. The data is indexed by its geographic coordinates. The database size parameters are summarized in Table 2. Both USGS and SPIN-2 data continue to arrive -- so these numbers will have increased by the time you read this.

| Table 2: As of June 24 1998, the TerraServer database has 718 GB of user data stored in 174 million records. About 200 GB of additional space is consumed by overhead (about 25%). The remaining space is used for indices, catalogs, recovery logs, and temporary storage for queries and utilities. The database has a formatted capacity of 2.2 TB. | | | |
|---|---|---|---|
| **Total Disk Capacity** | **Unprotected** | **after RAID5** | |
| | 324 disks x 9 GB = 2.9 TB | **4 x 595 GB volumes = 2.4 TB** | |
| **Database Size** | **Area** | **Bytes** | **Million Records** |
| **Gazetteer** | | .16 GB | 2.6 |
| **USGS** | 2.2 sq tera-meters @ 1m (JPEG) | 554 GB | 133 |
| **SPIN-2** | 1.3 sq tera meters @ 1.6m (TIFF) | 164 GB | 38 |
| **Total User Data** | **5 square tera meters** | **718GB** | **174** million records |
| **Overhead Space** | | 200 GB | |
| **Index, Catalog, log** | | 94 GB | |
| **Temp Space** | | 100 GB | |
| **Total DB size** | | **1.2 TB** | |

**SPIN-2 Theme:** The raw SPIN-2 data is divided into 4 40km x 40km photographs and scanned at 1.56-meter resolution. One 40km x 40km photo is picked to be the "anchor" photograph. Five points are goe-located in the "anchor" photograph. The anchor its 3 siblings are each quartered (a total of 16 images) and rotated the same angle. The rotation so North is up, optical distortion minimized, and geo-located pixels in the "anchor" image are accurate to 50-meters. The "sibling images" are geo-located such that pixels from the siblings can be aligned with the anchor image. The Sovinformsputnik and Aerial Images do all this work. The data is then sent to Microsoft on 20 GB DLT magnetic tapes. The typical image is 300 MB. It would take three years to download such an image over a 28.8 modem. We slice-and-dice these images into 10-KB tiles that can be downloaded in a few seconds. The slice and dice step produces four products:

*Jumps*: JPEG compressed images covering a 1 x 1.3 km area at 32-meter resolution
*Thumbnails*: JPEG compressed images covering a 1 x 1.3 km area at 16-meter resolution
*Browse*: JPEG compressed images covering a 1 x 1.3 km area at 8-meter resolution
*Tiles*: JPEG images that cover a 240 x 300 m area at 1.6-meter resolution.

The key property is that these tiles can be downloaded quickly over a voice-grade telephone line. The tile images are lightly encrypted. Each image, along with it's meta-data (time, place, instrument, etc.,... ) is stored in a database record. Each resolution is stored in a separate table. This data can be cross-correlated with the Gazetteer and other sources by using the Z-transform. In January 1998, we had 16 million tiles, and 650 thousand browse, thumbnail, and jump images. This totals 101 gigabytes of user data (compressed). It is 800 GB of uncompressed data. Loading continues as more data arrives from the Sovinformsputnik.

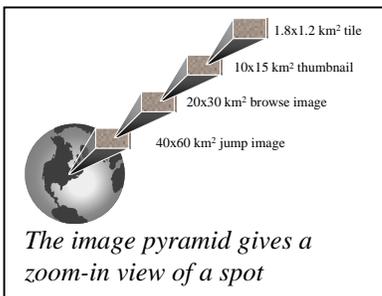

*The image pyramid gives a zoom-in view of a spot*

1.8x1.2 km² tile
10x15 km² thumbnail
20x30 km² browse image
40x60 km² jump image

**USGS Theme**: The USGS images are handled similarly. They arrive on DLTs from the USGS. The slice and dice step produces four products:

*Jumps*: JPEG compressed images covering a 1 x 1.3 km area at 32-meter resolution
*Thumbnails*: JPEG compressed images covering a 1 x 1.3 km area at 16-meter resolution

> *Browse*: JPEG compressed images covering a 1 x 1.3 km area at 8-meter resolution
> *Tiles*: JPEG images that cover a 150 x 225 m area at one meter resolution.

Each image, along with it's meta-data (time, place, instrument,  ) is stored in a database record. Each resolution is stored in a separate table. Today we have 96 million tiles and 1.5 million browse, thumbnail, and jump images. This totals about 800 gigabytes of user data. The US is about 9.8 million square kilometers, so this is about 30% of the US. Important areas have not yet been digitized. The USGS will provide additional data as it becomes available. They plan to digitize the entire country by the year 2002.

### Logical Database Design for Image Data

The images are stored in the SQL Server database along with their meta-data. The tiles, thumbs, browse, and jump images are kept as SQL *image* fields as part of relational records. The schema is shown in Figure 2.

**Meta-data**: All the original metadata for each large image is stored in the OriginalMetaData table. A user can ask the TerraServer for the lineage of a particular image. In that case, the TerraServer returns the appropriate record from this table. This data describes the data set in detail. The original meta-data table has about 100 fields. These fields describe the instrument, when the image was acquired, what format it is in, when and how it was processed, the resolution and size, and so on. The ImgSource field says USGS or SPIN2 now, and the image type is either JPEG or TIFF. Currently, all the TerraServer data is in JPEG3 format.

The ImageMeta table stores the meta-data for each jump-browse-thumb tile group. The meaning of most of the fields is obvious, the ImgStatus field allows us to hide sensitive places. Both the United States Government and the Russian Government want to be able to quickly hide a region in case there is a conflict. One can hide an area by updating this field. The TerraServer will not show images that have a "false" status.

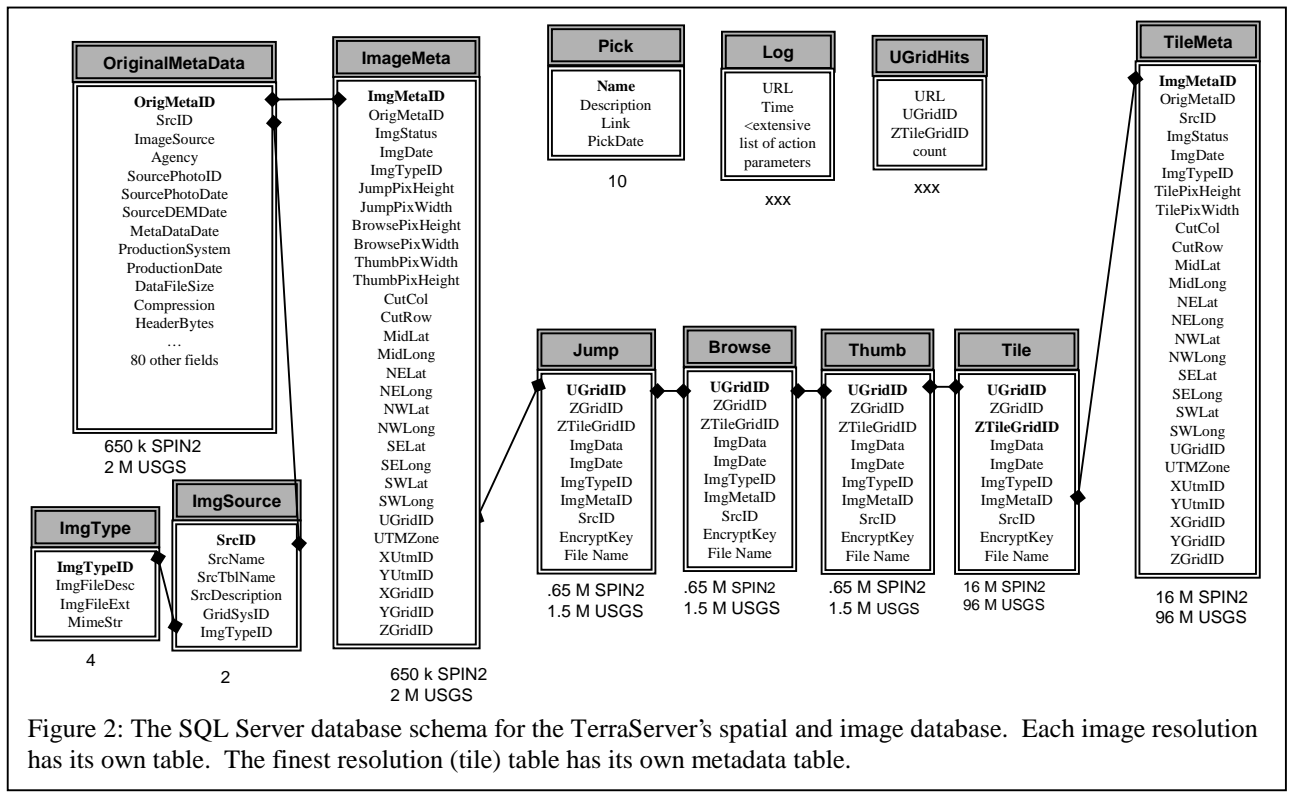

Figure 2: The SQL Server database schema for the TerraServer's spatial and image database. Each image resolution has its own table. The finest resolution (tile) table has its own metadata table.

**Jump, Browse, Thumb, Tile are separate tables:** Each of the jump, browse, and thumb image resolutions is kept in a separate table. We could have kept them all in one table, but the programming was more convenient if they were in separate tables.

**SPIN2 and USGS are separate tables:** Similarly, although the schemas are identical, we segregated the SPIN2 data from the USGS data. This was done just to simplify the programs -- a program wants either SPIN2 data or USGS data at any one time.  This simplified both programming and  index design.  For example, we require UgridId values be unique in the USGS tables because we only keep one image per UgridId.  Yes, that means we discard old USGS images as we get newer ones.  We do not keep any "Zgrid" indices on the USGS tables since all searches are by Ugrid.  Though we do compute a ZgridId so we can find a SPIN-2 image that overlays this image.  The reverse is true for SPIN-2 tables.  Here we keep Zgrid indices and do not maintain Ugrid indices.  We also keep duplicate ZgridIds for SPIN-2 data since we do want to show an older image and the latest image of the same spot.  Because we had two separate search schemes, two separate image retention schemes, and frankly, two separate application requirements because of the desires of two separate vendors, we decided to maintain two separate tables.  This makes it fairly easy to add a third data provider with an entirely new search scheme and application.

**The Pick table:** The Pick table is a list of recommended or interesting images.  Recent or topical locations can be added to this table and they will appear in an *Image Picks* recommendation web page.  The log and hit tables are used by the administrators to track how the system is used.  These tables record each request and record a count of requests for each grid-id and gazetteer entry.

**The Tile table:** The bulk of the database is in the tile table.  It contains the 110 million tiles for USGS and SPIN2.  Each tile has some meta-information giving its location in the grid system, the date the data was acquired, and then the 10 KB image itself.  The SPIN2 data tiles are lightly encrypted.  The encryption key is stored in the metadata.   A parallel TileMeta table stores additional metadata about each tile.

**Lookup by GridID:** The USGS and SPIN2 databases are typically accessed by spatial location.  The user points to some spot on the planet, and asks for the images around that spot at some specified resolution.   Thumbnail (16m/p) is the default resolution.  Suppose you live latitude 40ºN, longitude 140ºW.  This translates to a ZGridID and a UGridID.

Depending on the theme, the TerraServer will lookup images at or near that ZGridID (SPIN2) or UGridID (USGS).  These IDs are constructed using the Z-transform [Samet] so that nearby IDs are close to one another on the map.  Requests for adjacent images will probably ask for data that has already been pre-fetched by SQL into the database cache.   The use of GidIDs makes spatial lookups easy, and clusters nearby data together on disk.

### Database Design for Gazetteer

Using maps to navigate to images of particular areas is universal: it transcends language barriers, and it is intuitive. However, it often requires several steps to zoom in on a spot. Worse, if the viewer does not know the geography of a place, the graphical *point-and-see* metaphor may not be very useful. Someone who does not know the Washington DC area might have difficulty finding the White House or the Pentagon.

**Name lookup:** A name lookup application that quickly takes the user to a particular place solves this problem. We implemented an English-language name lookup system for the TerraServer. We began with Microsoft's Encarta™ World Atlas 97 gazetteer (also called VirtualGlobe™). Microsoft geographers, using many public and private sources, have been refining the Encarta gazetteer for several years. Today it has over 250 different countries, 1083 states, and 1,089,897 places. Many countries, states, and places have alternate names. Indeed, there are about 1448 country names, 3776 state names. There are at least 25 different ways to spell Albania, and 29 alternate names for the country of Yemen.

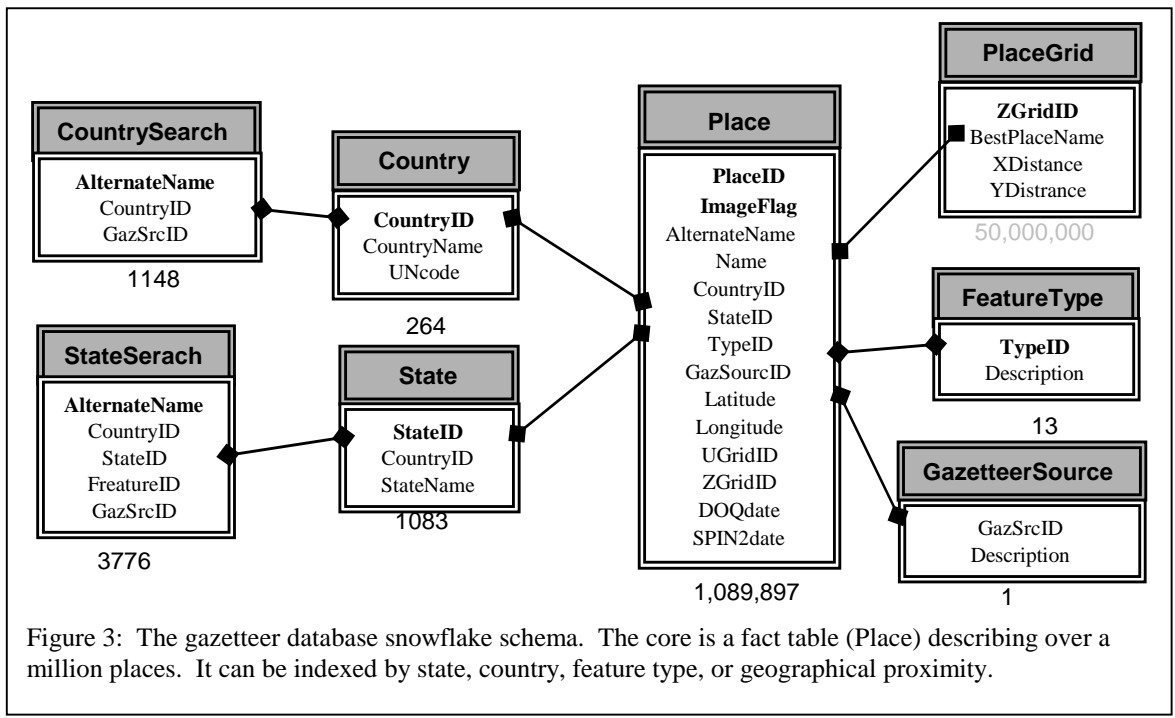

Figure 3: The gazetteer database snowflake schema. The core is a fact table (Place) describing over a million places. It can be indexed by state, country, feature type, or geographical proximity.

**Gazetteer is a snowflake-schema:** We extracted the Encarta gazetteer into the schema is shown in Figure 3. This is a classic snow-flake decision-support schema with the place-table at the core and the attribute tables radiating from the core.

The CountrySearch and StateSearch tables have the many alternate spellings for each country or state. For example, Uzbek S.S.R., Uzbek Soviet Socialist Republic, Uzbekskaya Sovetskaya Sotsialisticheskaya Respublika, Uzbekskaya SSR, Zbekiston Respublikasi and Uzbekistan all map to the nation of Uzgekistan.

**Gazetteer features:** The feature type table encodes the type of a place: (1) Airport/Railroad Station, (2) Bay/Gulf, (3)Cape/Peninsula, (4) City, (5)Hill/Mountain, (6) Island, (7)Lake, (8) Other

Land Feature, (9) Other Water Feature, (10) Park/Beach, (11) Point of Interest, (12 ) River.   The GazetteerSource tells what source the Place data record came from  (currently, Encarta is our only source).   The PlaceGrid table is not yet populated, but when it is, it will map each spot on the earth (each ZGrid ID) to the closest point for which the TerraServer has image data.

**Place table links to ZGrid and UGrid.**  The place table is the core of the schema.   It is heavily indexed by the other tables.  In addition the place table has 8 indices of it's own - six for the gazetteer lookup and two for UGrid and ZGrid lookups.   The place table is denormalized: we did not create a "PlaceSearch" table that factors out the *AlternateName to Name* mapping.  On average, there are only two names for a place, so a complete place record is stored for each instance.   This redundancy speeds access.   The meanings of most of the place table fields are obvious: they give the latitude, longitude, and grid ID of the place in both the USGS (Universal Transverse Mercator projection) and in TerraServer's SPIN2 ZGrid projection.  The place table record also has the place's latitude and longitude.   Each place table record has the most recent timestamp for USGS and SPIN2 images of the place, if the images exist.   A boolean flag (ImageFlag), indicates whether the place has an image.

**Indices to support arbitrary lookups:** The user interface allows users to specify any subset of place name (e.g. Paris or Pentagon), State (e.g. Wisconsin), country (e.g. France), or type (e.g. river).  If State is specified, but Country is not, we default to USA as the country. When searching for all places in a country, state, or type, the TerraServer shows the places that have images first in the list.  If the list is more than 10 elements, the search must continue just after the previous last record.  These requirements translate to a need for the following five indices in addition to the PlaceID, UgridID and ZgridID indices:

| Indices for Place Table | |
|---|---|
| akplace1 | ImgFlag, AlternateName, typeID |
| akplace2 | ImgFlag, countryID, stateID, AlternateName, typeID |
| akplace3 | ImgFlag, countryID, stateID, typeID |
| akplace4 | ImgFlag, countryID, AlternateName, typeID |
| akplace5 | ImgFlag, countryID, typeID |

**Showing places with images first:** The image flag causes all places with satellite images to sort low (early) in the answer set.    The rich indexing of the place table means that all the lookups, after cascading through the small snow-flake dimension tables, go directly to an index lookup of the place table.  The queries that access this table each select the first ten qualifying records.  To tell the optimizer that the query will not read the entire set, each query has the hint "FastFirstRow".

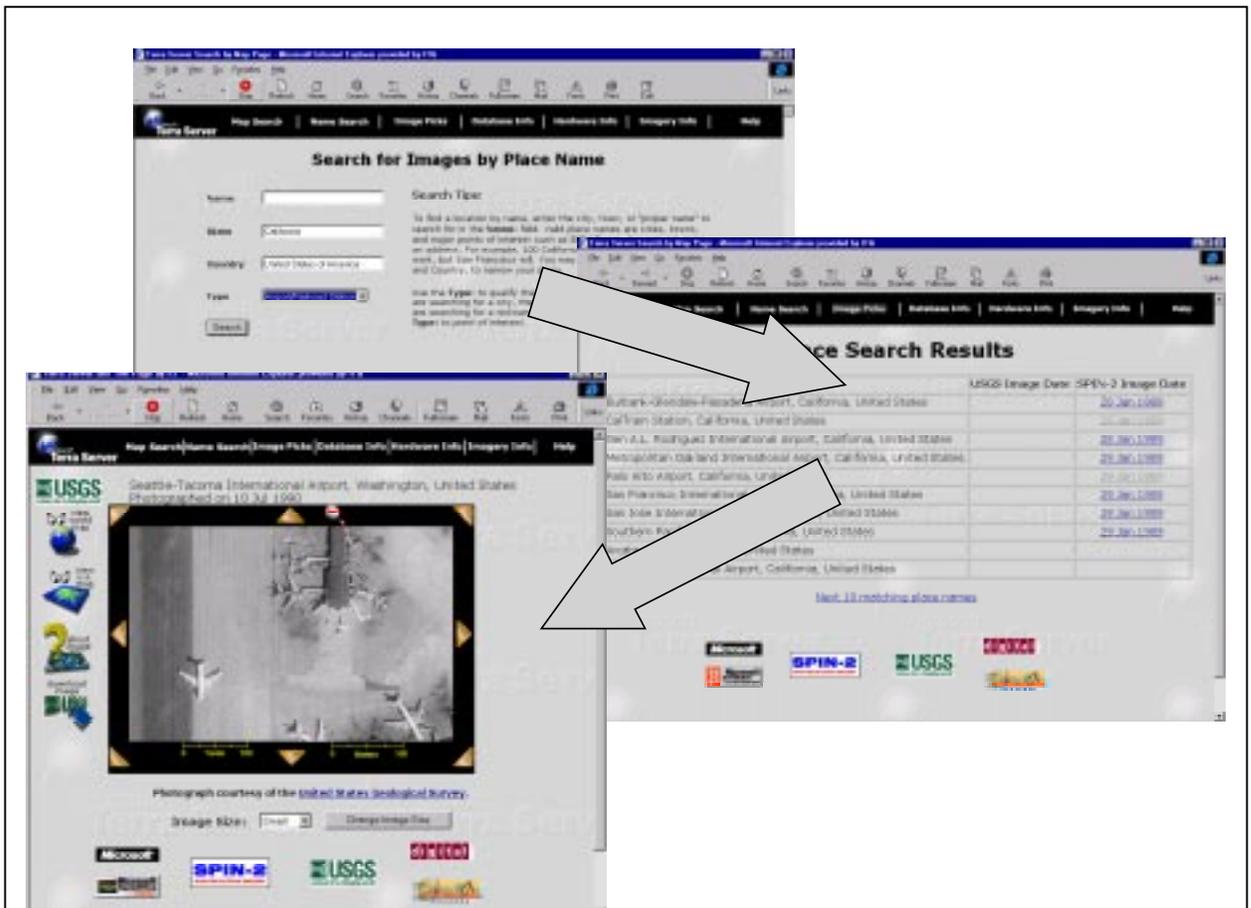

*Figure 4: An example of using the Encarta gazetteer to navigate the TerraServer by name. The user specifies the any combination of a city name, state, country, or place type. The gazetteer returns matching places and web links for all those places for which the TerraServer has image data.*

The name lookup process flows as follows:
1. The client fills in some fields of the HTML form shown in Figure 2. In this particular case, the client asked for airports in California. The form invokes and active server page on the TerraServer.
2. The active server page invokes an SQL stored procedure to find up to ten places matching the criteria.
3. There are ten possible queries, the stored procedure uses a case statement to pick the appropriate query (e.g.
    SELECT  ImageFlag, Place.AlternateName, USGSdate, SPIN2date
    FROM  ( (Place JOIN FeatureType ON Place.FeatureID = FeatureType. FeatureID )
        JOIN StateSearch ON Place.StateID= StateSearch.StateID)
        WHERE FeatureType.Description = 'Airport'
        AND    StateSearch.AlternateName = 'California'
        ORDER BY ImageFlag, Place.AlternateName)
        FASTFIRSTROW.
4. The query opens an SQL cursor, and fetches 10 qualifying rows. These rows are returned to the active server page, which formats the answer into an HTML page. This page contains the URLs for each matching place. Each such URL points to the TerraServer and includes a theme (USGS or SPIN2), and a grid-ID for that place. Given this information, the TerraServer can materialize an HTML page for that spot.

5.  The user can scroll through the pages that match his query (each page has up to ten matches). As the client scrolls forward, the SQL query opens the cursor where it left off, and reads the next ten qualifying places. If the user selects link and follows it, that sends a request to the TerraServer to materialize a image page for that place (in a USGS or SPIN2 theme.)

Navigating by name is one of the most popular ways to get oriented initially. Once the user has found a locale, zooming and panning are the most convenient ways to navigate.

**Gazetteer is used to show image place name**: The gazetteer provides another useful feature. When displaying an image, the TerraServer looks up the place name of the center of the image. So, for example, as one pans down the Washington Mall, the place names progress through the various monuments and museums. This is a valuable aid to helping uses orient themselves.

## Physical database design

**One SQL Database:** The TerraServer physical database design is very simple. The image and gazetteer data are all stored in one SQL Server database. SQL Server spreads the database and its recovery log across all the logical volumes. SQL Server also manages the use of physical memory for the buffer pool. Previous sections described the indices on the various tables.

**SQL Server file groups**: In SQL Server 7.0, a WindowsNT file is the basic unit of allocation, backup, and recovery. The SQL Server database is built from these files. SQL Server 7 spreads a table across a group of files. By default, all tables go into a default file group. This group stores the master database and other system information. We stored the gazetteer database in this default group, but defined a special file group to store the image data. This image file group contains all the image and tile files. A third group of files is dedicated to the database log.

**RAID 50 makes four 600GB disk volumes**: The TerraServer's physical database design is very simple. The database is mapped onto 4 huge disk volumes, each volume is almost 600 GB. Hardware RAID5 provided by Compaq StorageWorks converts the 324 disks into 28 large RAID5 disks. NT software striping (RAID0) is used to convert these 24 disks into four huge logical volumes. Therefore, SQL Server sees four huge disks.

The Microsoft TerraServer has been operating two DEC AlphaServer systems for a total of 18 months. One system has 324 4.3 GB drives the other 324 9.1 disk drives. In the 18 month period, we had seven disk drives fail – four 4.3 GB drives and three 9.1 GB drives. Never once did we loose any actual data. The lost disk was successfully rebuilt without any loss of data or any disruption in service to our applications.

**20 GB files are a convenient size**: To ease tape handling, and to provide some granularity of backup and recovery, we defined thirty 20 GB files on each disk. This file size fits nicely on current tapes. Actually, SQL Server and WindowsNT grow these files on demand as the database fills. Therefore, we just described the 120 (4x30) files to SQL as the file group.

**Its that simple:** That is all there is to say about the physical design. SQL Server and WindowsNT manage the huge disk array, they place the image data on the disks, index them, and retrieve them on demand.

### Data Loading

**Input was 300 DLT tapes:** The USGS sent us approximately 210 DLT tapes. Aerial Images sent us approximately 80 DLT tapes. These tapes contain over five terabytes of uncompressed images. The first step was reading these tapes to disk. Fortunately, the TerraServer was located in a demonstration area at Microsoft's Executive Briefing Center. There are many high-performance disk farms on display in this center. Consequently, we had access to almost five terabytes of disk space in order to process the images.

**The load process:** Conceptually, the process of loading data into TerraServer is very simple. Tapes arrive from the USGS and SPIN-2 containing uncompressed image files. The files contain too much data to be downloaded over the Internet and are not in a format recognized by web browsers. More importantly, the individual image files need to be modified so they adjacent images line up correctly when viewed on a web page.

**Image processing:** We wrote an image processing program that would take several large input files at a time, compute the image's location on earth, and where necessary, merge pixels from multiple files into one single photo. The "merged" images would "cut into smaller" images and compressed in the Jpeg file format. A single image covers a constant size on the ground. USGS images are 1800 by 1200 pixels. SPIN-2 images are $1/48^{th}$ of a degree of longitude by $1/96^{th}$ of a degree of latitude. These images were "sub-sampled" to create the image pyramid of 8m per pixel, 16m per pixel, and 32m per pixel. Full resolution USGS images were tiled 8 by 8 into images 225 pixels wide by 150 pixels tall. SPIN-2 images were tiled 5 by 5 and vary in their height and width.

We refer to the TerraServer sized images as "cuts". There are 67 image files and 134 meta data records generated for each USGS "cut" processed. There are 28 image files and 54 meta data records generated for each SPIN-2 "cut". This data is output to "flat files".

**Workflow system to manage image cutting:** Due to the intensive nature of preparing the files for loading into SQL, a workflow system was built using several applications to manage the cutting and loading process. This enabled many steps of the process to run in parallel. Each step of the process is recorded in the database, and an Active Server Page, or Web interface, is used to observe and manage the workflow.

**Load manager:** Microsoft also wrote a load manager application that feeds the data files into the database using the SQL Server 7.0 loader APIs (ODBC and BCP). Using several parallel streams, the data loads at approximately 2MBps. Although SQL is capable of a much higher load rate, the load process is constrained by the scan, slice, and dice process.

**Hardware for image processing, cutting, and loading:** Several Microsoft partners donated equipment for use in the TerraServer image cutting and loading process.

- Compaq provided two, three processor Alpha 4100s with 2GB of RAM each connected to 125GB (250GB total) of StorageWorks disk arrays.
- Clarion provided a 1TB storage array – 110 9.1 GB hard drives. 60 drives are connected via Fiber Channel to one a 4 processor 200 MHz 512MB RAM and the other 50 drives are connected via Fiber Channel to a second 4 processor 200 MHz 512 MB RAM.
- EMC provided a 500 GB Raid storage array. 250 GB is connected to a 4 processor 200 MHz 512mb RAM server and 250 GB is connected to a DEC/Intel PC with 2 processors and 512MB RAM.

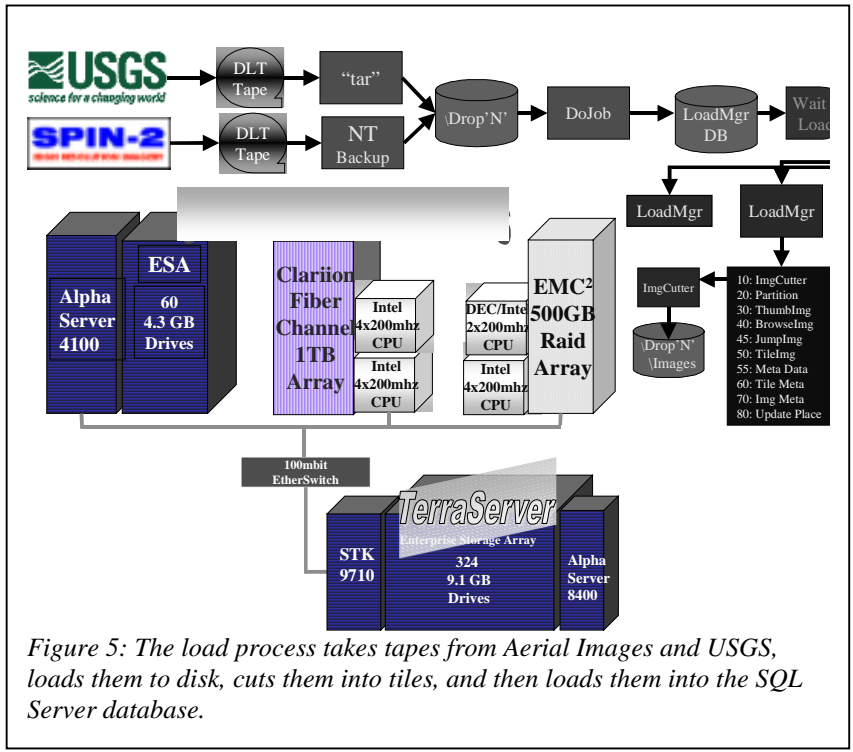

*Figure 5: The load process takes tapes from Aerial Images and USGS, loads them to disk, cuts them into tiles, and then loads them into the SQL Server database.*

These systems all ran the same complex of programs. Two or more Load Manager programs are launched on each CPU. Load Manager's fetch jobs for assigned to their node. Because the USGS image data naturally divides into "UTM zones", each server is assigend to a separate zone.

USGS jobs are fed to each machine in a "bottom to top of each zone" order. For example, node Fiber1 processed Zone 11 (Arizona, Nevada, Idaho, etc.). Jobs were submitted to it starting at the southern end of Arizona and moving towards the last job in northern Montana. The Load Manager monitors when the Image Cutter program completes "a set of rows than span across one zone". We called this a band. The US was divided into 100 bands.

When each band completes, the Load Manager inserts the meta data and image data into the database using the BCP (bulk copy program) API. For SQL 7.0, this API is based on ODBC. Once band was loaded, the Load Manager would ask for the next band to execute. We found that we achieved optimal performance with two Load Managers running on each node. One would always be running the Image Cutting program and the other Load Manager would be inserting data.

A third Load Manager monitors the progress made by the other two. When the files cut by the first two Load Manager have been loaded, the third Load Manager:
- deletes the compressed Jpeg files that have been loaded,
- invokes a Legato script to backup the full resolution cut images that were merged from two or more USGS DOQ photos,
- and deletes them disk.

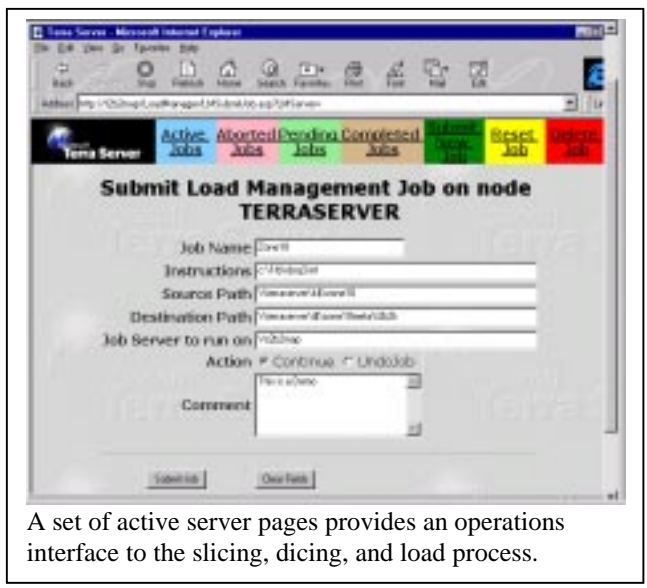

A set of active server pages provides an operations interface to the slicing, dicing, and load process.

A slightly different process is used for SPIN-2 images. Images are international. Photos taken from the same roll of film are grouped together. Pixels from one-to-four separate 300 MB SPIN-2 image files are merged and cut by the Image Cutter. Two LoadManager programs are invoked per each CPU. While one LoadManager was cutting data, the other was inserting compressed images and meta data into the database.

**STC TimberWolf tape robot**: We first processed the SPIN-2 imagery. Then we loaded the USGS images. In mid-January, we discovered a flaw in the registration of the SPIN-2 images and so had to reprocess them a second time. All this tape handling taught us the virtue of a reliable tape robot. It is also essential for unattended backup and restores. We used a Storage Technology 9710 TimberWolf tape robot. It can store up to 25 TB and has six DLT7000 tape transports. We sustained over 20 MBps transfers to and from it (peak rate was 80 MBps).

**Merging data from adjacent areas:** The SPIN2 images have been partitioned into 40x40 km patches. Tiles from different satellite image passes often overlap. To provide a seamless mosaic of an area, these patches must be combined into a large image, and then sliced and diced into the image pyramid described earlier. Consequently, all the patches for an area need to be online when the SPIN2 data is being processed. Similarly, the USGS data for a particular zone is spread across all the tapes the USGS sent us. We had to read almost all the tapes to build an image of a zone. To minimize tape operations, we decided to process all the SPIN2 data and then all the USGS data online.

**Automated slice&dice process**: Once the data is online, slicing and dicing it into tiles, thumb, browse, and jump images is compute intensive. This step produces a huge number of files (20 million for some UTM zones). These files are then copied into the database. We built a simple workflow system to manage the cutting and loading process. Many steps can run in parallel. Each step is restartable. Each step is recorded in a database. An active server page (web interface) is used to observe and manage this workflow.

**Automated Database Load process:** Once the files have been created for an area, the SQL Loader is used to insert the images into the database. The loader program accesses the data in Zgrid order and then inserts the images along with their metadata into the SQL database. SQL Server now allows parallel database loads into the same table, so several streams can run in parallel. We can load at a rate of about five megabytes per second. Locally, SQL Server can load data at 15MBps, but the load program could not produce data that quickly. Opening the many files and passing them to SQL limited the load program to a peak of 2 MBps.

**Adding image data adds metadata and updates gazetteer**: The database load also populates the meta-data tables and updates the gazetteer database to record the most recent SPIN2 and USGS data.

**Incremental load:** Now that the TerraServer is operating, data is being added by inserting data rather than by doing bulk data load. The data rate is lower (one MBps) but still adequate for our needs.

## Backup & Restore

We archived the data up to tape after each processing step. Thus far, we have not had to use those backups. The backup units are WindowsNT files. We use the Legato NetWorker 4.4 backup utility to manage this work.

**SQL Server Incremental backups:** In addition, while the system is operating we take online tape backups of the database. SQL Server 7.0 supports incremental backup (only changed pages are archived). Since the TerraServer is an insert-mostly database, these backup tapes are comparable in size to the amount of data that has been loaded since the last backup.

Table 3 shows the data rates we achieved. At these rates, we can reconstruct the entire TerraServer in a day, and can recover a damaged file in an hour.

| Table 3: Backup and Restore data rates. | |
| --- | --- |
| Read a file from tape | 200 GB per hour |
| Backup a file to tape | 200 GB per hour |
| Archive the whole TerraServer Database | 6 hours |
| Online Archive the TerraServer after a 10 GB insert | 1 hour |
| Restore a SQL Server file (media time). | 10 minutes |

We found the StorageTek hardware and Legato software to be very reliable. Actually, no one cares about backup – you only care about restore. We've backed up the TerraServer database and restored it several times on different systems and using different StorageTek robots. Not once have we experienced a tape error or an usable database.

Our goal was to be able to backup the entire 1TB database in less than 8 hours. Our StorageTek TimberWolf 9710, which has 10 DLT drives connected via 5 Fast Wide Differential SCSI channels to the DEC Alpha 8400 backs up our Terabyte SQL database within six hours.

StorageTek offers a wide range of automated tape "robot" subsystems ranging from 2 drive, 100 tape slot 3.5TB robot to a huge, 154 TB tape robot sub-systems. Thus, StorageTek can scale as TerraServer database scales. StorageTek supports several tape media types – Digital Linear Tape (DLT), IBM 3480, and other popular media. Today Microsoft TerraServer uses DLT, a common media type in the image-processing world, but we want to be able to change to other media types that emerge in the future.

**Legato NetWorker is simple to run.** Legato has Windows based admin interface. A graphical display of our TimberWolf 9710 tape robots physical structure and our disk system allows us to simply "drag and drop" files to a tape device to schedule a backup. More sophisticated options are available in dialog windows to control scheduling and placement of data on media.

In addition to backing up and restoring our database, we also receive tapes from the USGS and SPIN-2 in a variety of tape formats. Using Legato's Admin interface, we can control the physical loading of tape media into inventory slots, load into tape drives, tape ejection to tape slots, and removal from the system all in a graphical user interface.

**Legato integrates with SQL Server backup.** In addition to backing up our database files, Legato NetWorker integrates with SQL Server's Backup utility. This allows us to backup our Terabyte database while users are accessing the database.

## Microsoft TerraServer Hardware

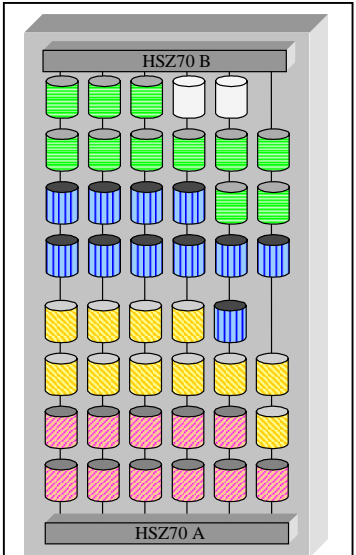

*The Configuration of the StorageWorks disk array: each RAID controller manages two RAID5 arrays of 11 disks to make an 85GB RAID5 disk. Each controller has a 64 MB cache. There are two spare disks.*

**Processors**: TerraServer runs on a Compaq Alpha 8400 system with 10 GB of memory. This system has eight 440 MHz Compaq Alpha processors. The 8400 can support up to 160 PCI slots. In our configuration, we host seven KZPBA dual-ported Ultra SCSI host bus adapters -- one for each of the seven disk storage cabinets. There are also six KZPSA SCSI host bus adapters for back up and for boot disks.

**Disks**: The system has seven storage cabinets each holding 46 9 GB drives, for a total of 324 drives. Their total capacity is 2.9 terabytes. The drives are configured as RAID5 sets using 14 HSZ70 RAID controllers each with 64MB of memory. Each pair of controllers manages 46 disks on six SCSI strings. Each controller manages two RAID5 sets of 11 disks. That leaves two spare disks in case a disk fails. So the 14 HSZ70s collectively manage 28 RAID5 sets, each storing 85 GB. Windows NT file striping (RAID0) is used to collapse these 28 RAID5 disks into four logical volumes. Each logical volume is 595GB (= 7 * 85GB). The resulting four logical drives are each given a drive letter.

SQL Server stripes the database across these 4 logical drives. This mapping is dictated by the fault-tolerance properties of the StorageWorks array. The design masks any single disk fault, masks many string failures, and masks some controller failures. Spare drives are configured to help availability. In the end, SQL Server has just over 2.4 TB of RAID5 protected storage. The StorageWorks array has been trouble-free.

**Tapes**: The TerraServer has a six-station STC 9710 tape robot with a near-line capacity of over 5 terabytes. This robot is used for backup and recovery. It is also used to import data from other sources.

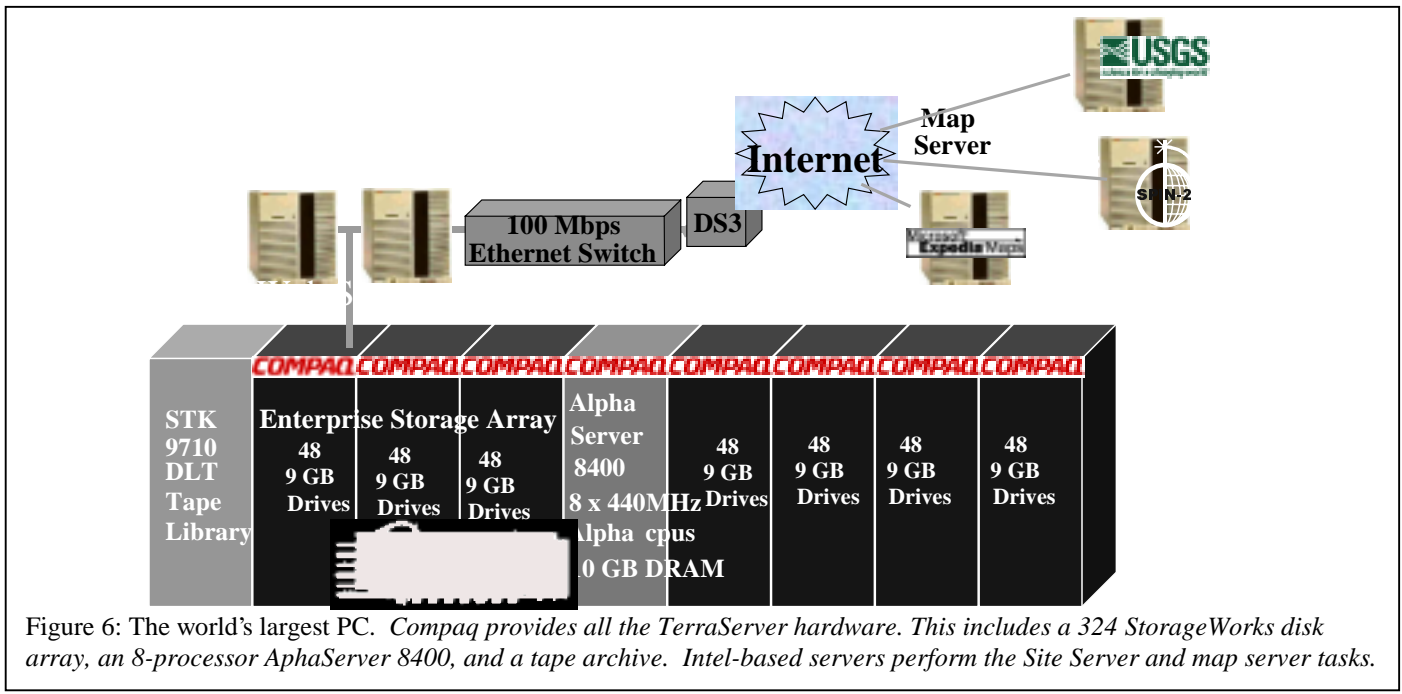

Figure 6: The world's largest PC. *Compaq provides all the TerraServer hardware. This includes a 324 StorageWorks disk array, an 8-processor AphaServer 8400, and a tape archive. Intel-based servers perform the Site Server and map server tasks.*

**Map Server**: TerraServer accesses the Expedia Map Servers. These servers are a shared resource available to Microsoft Internet applications. TerraServer shares four map servers with Microsoft Expedia Travel Services, Microsoft Sidewalk™ Cityguide, MSN's CarPoint™ automotive service, and MSNBC. The servers run on dedicated Compaq servers (Intel processors) each with four processors and 256 MB of memory.

**Site Server Commerce Edition:** Aerial Images and the USGS each host a E-commerce web site tailored to interface with Microsoft TerraServer. Both sites run Microsoft Site Server Enterprise Edition V3.0 software on Windows NT Server 4.0, SQL Server 6.5 and Internet Information Server 3.0. Aerial Images chose a Compaq Alpha 800 server cluster with 100 GB of disk. Aerial Images maintains the server in Raleigh, N.C. The USGS has chosen a multi-processor Intel based server from Gateway 2000. The USGS houses the server at the USGS EROS Data Center in Sioux Falls, South Dakota.

**Networking**: The TerraServer is behind the Microsoft firewall. It is connected via dual 100 Mbps Ethernets to high-speed Internet ports.

**Slicing and Dicing**: Most of the Slicing and Dicing of the images was done on the TerraServer with the aid of two Compaq 4100's with 200GB of disk storage and some 4-way Intel processors. The load processes deliver new data from this node to the TerraServer over a switched 100 Mbps Ethernet. Image processing by the Sovinformsputnik, Aerial Images, and the University of California at Santa Barbara, used Windows NT Server on Intel systems.

# Assessment

The TerraServer is a simple application, but it involves many tools: HTML, Java, VBscript, HTTP, IIS, ASP, OLE DB, ODBC, and SQL Server. As such, it is a showcase web application. Once the rough design was chosen, it was fairly easy to design and configure the database.

We were novices at the many data formats used in geo-spatial data — but we learned quickly as we did the slicing and dicing. Working with our geo-spatial-data mentors at Aerial Images, USGS, and the UCSB Alexandria Digital Library vastly accelerated this process.

Once the first user-interface was built, it was clear that we needed a better one. We are now on the forth iteration of that design. The design has converged, but designing intuitive user interfaces is one of the most difficult aspects of any system.

Once we understood the process, the slicing, dicing, and loading went very smoothly -- but it was the bottleneck in bringing the real TerraServer online. It takes a huge amount of computation to process five terabytes of image data. Although we were using Alpha and Beta versions of the next SQL Server, it gave us very few problems. The Compaq Alpha and StorageWorks equipment performed flawlessly. Having great tools makes it possible to experiment.

The SQL Server, IIS, and Windows NT Server management tools were a big asset. Overall, the project was relatively easy. One area that gave us difficulty was tape backup and restore speeds. That has improved dramatically since we started. We can now backup at a rate of 80 GB per hour.

Windows NT 5.0 Enterprise Edition supports 64-bit addressing on the Compaq Alpha processors. SQL Server V7.0 Enterprise Edition has been modified to exploit the power of very large memory (VLM). Although likely to improve Terraserver performance these technologies are not yet incorporated into the Terraserver.

TerraServer shows that Windows NT and Microsoft SQL Server can support huge databases on a single node. If one wanted to store an atlas of the entire landmass of the planet, it would be 25 times larger. Clearly, one would use larger disks and would have to use a cluster of 20 of these huge nodes in a design similar to the billion-transactions-per-day cluster.

## Other Scalability Projects

Concurrent with the TerraServer effort, other Microsoft groups undertook scale-up projects. This is a brief synopsis of those efforts.

**Billion Transactions per Day:** A 45-node Compaq Proliant cluster of 140 processors and 2.5 TB of disks was configured to run a billion transactions per day. The transaction profile was the DebitCredit transaction. The system had a classic three-tier architecture. The 20 front-end nodes submitted transactions via DCOM calls to Microsoft Transaction Server. MTS passe the transactions to 20 SQL servers. Approximately 15% of the transactions involved account transfers between two nodes. These distributed transactions were coordinated by Microsoft Distributed Transaction Coordinator. The system is currently running about 1.4 billion transactions each day in a demonstration set up on the Microsoft campus in cooperation with Compaq.

**100 M web hits per day**: A single dual processor Windows NT Server running Internet Information Server was configured with 1,000 virtual roots (web sites) all served by one web server. Interestingly, each web site models a large city and uses imagery from TerraServer as a graphic for the root page. Clients send HTTP requests to drive this web server. The server sustains an average rate of one hundred million -web hits per day. That is twice the hit rate of the www.microsoft.com site. This shows a breakthrough in web-server performance derived from the combination of Windows NT Sever and its imbedded Internet Information Server (IIS).

**50 GB Mail Server**: Most mail servers have a modest limit on the size of the message store. Microsoft Exchange Server had a limit of 16 GB. That limit has been extended to several terabytes. A server with a 50 GB message store was built (by filling it with Internet news groups).

**50,000 POP3 Mail users.** Microsoft Exchange was also stress tested with 50,000 clients sending 5 message per day, receiving 15, and checking mail twice an hour using the standard POP3 protocol. This ran on a single 4-processor node.

**64-bit Windows NT and SQL Server**: Windows NT 5.0 supports 64-bit addressing on the Compaq Alpha and Intel Merced processors. SQL Server has been modified to exploit the power of massive memory. In particular, a data analysis scenario using SAS running on SQL server completes within 45 seconds using the Compaq Alpha 64-bit addressing, but takes 13 minutes to complete if the system does not exploit massive main memory.

**High Availability SQL and Windows NT**: Fault-tolerance becomes very important when a cluster grows to have thousands of disks, hundreds of processors and thousands of applications. Windows NT Enterprise edition adds high-availability support to Windows NT 4.0. Most BackOffice applications have been made cluster-aware so that they can failover from one node to another if the node fails. To demonstrate this, Microsoft worked closely with SAP to make a high-availability version of R/3. With this version running on two nodes, one node does the R/3 front-end work, while the second node does the back-end SQL work (a classic three-tier application). If one node fails, the other node performs both the R3 and the SQL tasks. Users can continue using the R/3 system, even while the failed node is repaired and brought online. One-minute application failover times are common now.

# Summary

**Commodity-scalable servers have arrived**.  These scaleablity demonstrations show that with proper design, Windows NT Server, Microsoft SQL Server, and the other Microsoft BackOffice products can be used to solve the most demanding problems.  They demonstrate the key properties one wants of a scaleable system:
- Scalability — growth without limits.
- Manageability — as easy to manage as a single system, self-tuning.
- Programmability — easy to build applications.
- Availability — tolerates hardware and software faults.
- Affordability — built from commodity hardware and commodity software components.

The TerraServer demonstrates the extraordinary performance obtainable with commodity software components.  Commodity components give these systems excellent price-performance.  They are a breakthrough in the cost of doing business.  The cost of serving a page onto the Internet, delivering a mail message, or transacting a bank deposit has gone nearly to zero — it costs a **micro-dollar per transaction.**  One could use advertising to pay for such transactions — an advertisement pays a thousand times more per impression.

These applications were initially built by a few people in a few weeks.   The TerraServer design keeps evolving.  It is modular, so components are easily rewritten and plugged into the whole.  The TerraServer demonstrates the incredible power of the new tools created by the Windows platform and by the Internet.  The resulting applications have easy-to-use management interfaces, and have the manageability and availability properties essential to operating them.

Microsoft learned a great deal in building the TerraServer.  We fixed many performance bugs and eliminated many system limits.  There is still more to do to make these products even more self-tuning and self-managing.  That process is in full swing now.  Future releases of Windows NT Server and Microsoft BackOffice products will reflect these improvements.

# Acknowledgments

The TerraServer project is a consortium of four major participants – Microsoft, Compaq, Aerial Images, and the USGS. Each organization committed software, personnel, equipment, and intellectual property. Numerous other organizations have made substantial donations in equipment and expertise. Without their support, the TerraServer project would not be possible. These organizations in alphabetical order are:

**Aerial Images and Sovinformsputnik:** Aerial Images is a partner of Sovinformsputnik and is the sole distributor of Sovinformsputnik's SPIN-2™ imagery. Aerial Images and Sovinformsputnik have provided 350GB of SPIN-2 imagery for use on the TerraServer project. The SPIN-2 imagery provides the world-wide coverage for TerraServer. Approximately 50% of the SPIN-2 imagery covers Europe, China, Australia, and other non-U.S. locations. The Aerial Images partners gave this project enthusiastic support. Mr. Mikhail Fromtchenko and Dr. Victor Lavrov of Sovinformsputnik contributed their expertise and contributed the Russian imagery to the project

**Clarion, a subsidiary of Data General:** loaned us 1 TB of high speed disk storage. The Clarion disk array is connected via Fiber Channel interconnect to two WindowsNT servers. Fiber Channel is very simple to cable together and is very fast.

**Compaq Corporation:** provides the DEC Alpha 8400 and DEC StorageWorks ESA1000 complex of 2.4 TB of UltraSCSI disk capacity. The SQL Server 7.0 Database runs on the DEC 8400 and maintains the 1 TB SQL database of meta-data and imagery. In addition, DEC has provided a DEC Alpha 4100 with 1.3 TB of FWD SCSI disks, a pair of DEC Alpha 4100 systems with 250 GB of disk capacity, and DEC Prioris 6200 Intel based processor. These other systems are used to edit and load the imagery received from Aerial Images and the USGS into the DEC 8400 system.

**EMC Corporation:** loaned 500 GB of FWD SCSI disk storage for TerraServer's use. The EMC disk array held 750GB of FWD SCSI disks. 500GB was connected to a single Intel POCA 4-PentiumPro server.

**Intel Corporation:** loaned six "POCA" servers for use as image cutting and TerraServer demonstration equipment for use by Microsoft and Aerial Images. A POCA server is a four PentiumII processor system with 256 MB of memory. They have been fantastic workhorses while editing 4.0 TBs of raw imagery. To their credit, we had zero failures will all our POCA servers.

**Legato Networker:** The TerraServer database is backed and recovered by Legato Networker. The TerraServer project uses the Networker product on Compaq AlphaServersand Compaq Intel Servers. We chose Legato Networker as our primary backup/restore technology because (a) it was available on both Compaq Alpha and Intel Windows NT, and (b) it supported the Compaq TL894 and StorageTek 9710 tape robot at the time we started the TerraServer load in November 1997.

**Microsoft Research:** Microsoft Scalable Servers Research group based in San Francisco, California conceived the TerraServer application and has led the development of the custom software. All custom software has been developed using Microsoft development tools – Visual Studio 5.0, SQL Server 7.0 ,Internet Information Server 3.0 with Active Server Pages, Visual Basic Scripting Edition, and Microsoft Commerce Server, a component of Microsoft Site Server Enterprise Edition.

**Microsoft SQL Server Group:** Supported the TerraServer development with both human and financial resources.

**Microsoft Geo Business Unit:** Steve Smyth helped us with the Encarta Virtual Globe Gazetteer, and Microsoft's Geo Business Unit built the Java-based map control for us.

**Seagate Software's Backup Exec:** Aerial Images delivers 20km X 20km SPIN-2 imagery on CompacTapeIV media written using Backup Exec 6.11. TerraServer processes these tapes on Backup Exec running on DEC Alpha and Intel based Windows NT Servers.
**Storage Technology Corporation:** Provided an STK 9710 tape storage system. The 9710 contains six DLT7000 tape drives and 100 tape slots for a total capacity of 4.0 TB of tape. The STK is connected to the DEC Alpha 8400 using FWD SCSI connections. A single KZPSA FWD SCSI controller was connected to two DLT7000 tape drives per StorageTek.
**United States Geological Survey (USGS):** Provided 3.0 TB of Digital Ortho Quadrangle photographs. In addition, the USGS has provided in valuable assistance in geography and image processing of orthophotographs. The USGS also provided the source data for the Coverage Map application. The USGS' goal in the TerraServer project is evaluate the use of the Internet as a data presentation mechanism for the general public.
**University of California at Santa Barbara's Alexandria Digital Library project:** Jim Frew, Terry Smith, and others at UC Santa Barbara helped us slice and dice the USGS imagery. UCSB has a long track record of combining the earth and computer sciences into useful research.
was done jointly with Compaq Corporation and Aerial Images.

The Alpha Server™ division and the Storage Works™ division were especially helpful. The 324 StorageWorks disks on the TerraServer worked on day one and continue to work without event.
.